
\input harvmac
\input tables

\overfullrule=0pt
 
\def\A{{\scriptscriptstyle A}}

\def\Q{{\scriptscriptstyle Q}}
\def\R{{\scriptscriptstyle R}}

\def\T{{\scriptscriptstyle T}}

\def\V{{\scriptscriptstyle V}}

\def\Y{{\scriptscriptstyle Y}}


\def\CN{{\cal N}}

 
\def\a{\alpha}

\def\d{\delta}
\def\e{\epsilon}

\def\th{\theta}


\def\fifth{{1 \over 5}}

\def\half{{1 \over 2}}
\def\quarter{{1 \over 4}}

\def\third{{1 \over 3}}


\def\bar#1{\overline{#1}}
\def\bzero{{b_0}}
\def\ccdot{\hbox{\kern-.1em$\cdot$\kern-.1em}}

\def\Diag{{\rm Diag}}

\def\gtap{\raise.3ex\hbox{$>$\kern-.75em\lower1ex\hbox{$\sim$}}}

\def\Lambdab{{\Lambda^\bzero}}

\def\ltap{\raise.3ex\hbox{$<$\kern-.75em\lower1ex\hbox{$\sim$}}}
\def\Nc{N_c}
\def\Nf{N_f}
\def\R{{\scriptscriptstyle R}}

\def\sigone{\sigma_1}
\def\sigtwo{\sigma_2}
\def\sigthree{\sigma_3}
\def\sp{\>\>}

\def\therefore{{\hbox{..}\kern-.43em \raise.5ex \hbox{.}}\>\>}

\def\Vslash{V\hskip-0.75 em / \hskip+0.30 em}
\def\Wdyn{W_{\rm dyn}}

\newdimen\pmboffset
\pmboffset 0.022em
\def\oldpmb#1{\setbox0=\hbox{#1}%
 \copy0\kern-\wd0
 \kern\pmboffset\raise 1.732\pmboffset\copy0\kern-\wd0
 \kern\pmboffset\box0}
\def\pmb#1{\mathchoice{\oldpmb{$\displaystyle#1$}}{\oldpmb{$\textstyle#1$}}
      {\oldpmb{$\scriptstyle#1$}}{\oldpmb{$\scriptscriptstyle#1$}}}


\nref\ISTASI{K. Intriligator and N. Seiberg, Nucl. Phys. Proc. Suppl. 
 {\bf 45BC} (1996) 1, and references therein.}
\nref\SeibergI{N. Seiberg, Phys. Rev. {\bf D49} (1994) 6857.}
\nref\SeibergII{N. Seiberg, Nucl. Phys. {\bf B435} (1995) 129.}
\nref\Pouliotseminar{P. Pouliot, ``Chirality versus Duality'', Caltech theory 
seminar, September, 1996.}
\nref\Strassler{M. Strassler, Private communication.}
\nref\ILS{K. Intriligator, R.G. Leigh and N. Seiberg, Phys. Rev. {\bf D50}
 (1994) 1092.}
\nref\Elashvili{A.G. Elashvili, Funk. Anal. Pril. {\bf 6} (1972) 51.}
\nref\Georgi{H. Georgi, {\it Lie Algebras in Particle Physics}, 
 (Benjamin Cummings, Menlo Park) 1982.}
\nref\Wilczek{F. Wilczek and A. Zee, Phys. Rev. {\bf D25} (1982), 553.}
\nref\Harvey{J. Harvey, D.B. Reiss and P. Ramond, Nucl. Phys. {\bf B199} 
 (1982) 223.}
\nref\Intriligator{K. Intriligator, Phys. Lett. {\bf B336} (1994) 409.}
\nref\ChoKraus{P. Cho and P. Kraus, Phys. Rev. {\bf D54} (1996) 7640.}
\nref\CSSone{C. Cs\'aki, W. Skiba and M. Schmaltz, hep-th 9607210 (1996),
 unpublished.}
\nref\CSStwo{C. Cs\'aki, W. Skiba and M. Schmaltz, hep-th 9612207 (1996), 
 unpublished.}
\nref\IT{K. Intriligator and S. Thomas, Nucl. Phys. {\bf B473} (1996) 121.}
\nref\Pouliot{P. Pouliot, Phys. Lett. {\bf B359} (1995) 108.}
\nref\PouliotStrasslerI{P. Pouliot and M. Strassler, Phys. Lett. {\bf B370} 
 (1996) 76.}
\nref\PouliotStrasslerII{P. Pouliot and M. Strassler, Phys. Lett. {\bf B375} 
 (1996) 175.}
\nref\Kawano{T. Kawano, Prog. Theor. Phys. {\bf 95} (1996) 963.}
\nref\ADS{I. Affleck, M. Dine and N. Seiberg, Nucl. Phys. {\bf B256} (1985) 
 557.}


\def\LongTitle#1#2#3{\nopagenumbers\abstractfont
\hsize=\hstitle\rightline{#1}
\vskip 0.5in\centerline{\titlefont #2} \centerline{\titlefont #3}
\abstractfont\vskip .3in\pageno=0}
 
\LongTitle{HUTP-96/A060}
{Exact Results in SO(11) SUSY Gauge Theories}
{with Spinor and Vector Matter}

\centerline{
  Peter Cho\footnote{$^*$}{Research supported in part by the National Science 
Foundation under Grant \#PHY-9218167.}}
\centerline{Lyman Laboratory}
\centerline{Harvard University}
\centerline{Cambridge, MA  02138}

\vskip 0.3in
\centerline{\bf Abstract}
\bigskip

	We investigate the confining phase vacuum structure of supersymmetric 
$SO(11)$ gauge theories with one spinor matter field and $\Nf \le 6$ vectors.  
We describe several useful tricks and tools that facilitate the analysis of 
these chiral models and many other theories of similar type.  The forms of 
the $\Nf=5$ and $\Nf=6$ quantum moduli spaces are deduced by requiring that 
they reproduce known results for $SU(5)$ SUSY QCD along the spinor flat 
direction.  After adding mass terms for vector fields and integrating out 
heavy degrees of freedom, we also determine the dynamically generated 
superpotentials in the $\Nf \le 4$ quantum theories.  We close with some 
remarks regarding magnetic duals to the $\Nf \ge 7$ electric $SO(11)$ 
theories.

\Date{1/97}

\newsec{Introduction}

	During the past few years, remarkable progress has been made in 
understanding nonperturbative aspects of $\CN=1$ supersymmetric gauge 
theories.  Pioneering work in this area by Seiberg and collaborators has shed 
light upon such interesting strong interaction phenomena as phase transitions, 
confinement, and chiral symmetry breaking \ISTASI.  Their studies have also 
opened up several new directions for model building which potentially have 
important phenomenological applications.  Supersymmetric model investigations 
have thus yielded valuable insights into several basic issues in quantum 
field theory and particle physics.

	Many of the recent key advances were developed within the context 
of SUSY QCD which represents the prototype $\CN=1$ gauge theory
\refs{\SeibergI,\SeibergII}.  Unfortunately, it has often proven difficult to 
extend the new ideas beyond this relatively simple model to more complicated 
theories.  For example, finding weakly coupled duals to strongly coupled 
models with no simplifying tree level superpotentials remains an outstanding 
challenge despite significant theoretical efforts to uncover patterns among 
known dual pairs.  Confining phase analyses are generally more tractable 
than those which focus upon questions related to free magnetic and nonabelian 
Coulomb phases in various theories.  But even addressing confinement 
issues in models with more complicated matter contents than those like SUSY 
QCD with fields in only fundamental representations frequently requires one to 
overcome nontrivial technical problems.

	In this note, we investigate the confining phase vacuum structure of 
a supersymmetric theory based upon an $SO(11)$ gauge group with one spinor 
field and $\Nf \le 6$ vectors.  Our motivations for studying and presenting 
results on this particular model are threefold.  Firstly, we wish to describe 
a number of useful tools that greatly facilitate the analysis of this 
nontrivial theory's confining phase.  These tricks can be applied to the 
study of many other supersymmetric theories' low energy dynamics.  While most 
of the simple methods which we employ have been known to nonperturbative SUSY 
model experts for some time \refs{\Pouliotseminar,\Strassler}, we believe it 
is worthwhile to discuss these previously undocumented techniques so as to 
make them accessible to a larger community.  Secondly, the 32-dimensional 
spinor irrep of $SO(11)$ is pseudoreal.  Since no mass term for it can be 
written down, our model is chiral.  It may therefore have interesting 
applications for dynamical supersymmetry breaking.  Finally and most 
importantly, understanding the confining phases of $\CN=1$ theories is 
invaluable in searching for duals.  As we shall see, a dual to this $SO(11)$ 
model would act as a generator for several other magnetic descriptions of 
various electric theories.  While we have not yet constructed such a dual, our 
present analysis restricts its possible form.

	Our article is organized as follows.  In section~2, we discuss the 
low energy description of the microscopic $SO(11)$ model and identify gauge 
invariant operators which label its flat directions.  We demonstrate 
that this theory confines when it contains $\Nf \le 6$ vector fields.
We then proceed in section~3 to analyze the quantum moduli spaces in the 
$\Nf=5$ and $\Nf=6$ theories.  After adding mass terms for vector 
fields and systematically integrating them out, we also deduce the 
dynamically generated superpotentials in the $\Nf \le 4$ quantum theories.   
Finally, we close in section~4 with some remarks and speculations regarding 
duals to the $SO(11)$ models with $7 \le \Nf \le 22$ vector fields.

\newsec{The $\pmb{SO(11)}$ model}

	We begin our study of the $SO(11)$ model by listing its full symmetry 
group 
\eqn\anomsymgroup{
G = SO(11)_{\rm local} \times \bigl[ SU(\Nf) \times U(1)_\V \times U(1)_\Q 
\times U(1)_\R \bigr]_{\rm global},}
superfield matter content 
\eqn\anommatter{\eqalign{
V^{\mu i} & \sim \bigl( 11; \Nf ; 1,0,0 \bigr) \cr
Q^\A & \sim \bigl( 32; 1 ; 0,1,0 \bigr) \cr
\Lambdab & \sim \bigl( 1; 1; 2 \Nf; 8, 10 - 2 \Nf \bigr)}}
and one-loop Wilsonian beta function coefficient
\foot{We adopt the $SO(11)$ irrep index values $K(11)=2$, $K(55)=18$ and 
$K(32)=8$.}
\eqn\betafunc{b_0 = \half \bigl[ 3 K({\rm Adj}) -
\sum_{\rm{\buildrel matter \over {\scriptscriptstyle reps \>\rho}}} K(\rho)
\bigr] = 23 - \Nf.}
In the absence of any tree-level superpotential, the classical theory remains 
invariant under an arbitrary $G$ transformation.  But in the quantum theory, 
each of the $U(1)$ factors in eqn.~\anomsymgroup\ is anomalous.  The theta 
parameter in the $SO(11)$ Lagrangian undergoes a shift 
\eqn\shift{\th_{SO(11)} \to \th_{SO(11)} + C \a}
when an anomalous $U(1)$ rotation through angle $\a$ is performed.  As a
result, the spurion field 
\eqn\spurion{ \Bigl( {\Lambda \over \mu} \Bigr)^\bzero \equiv \exp \Bigl[ 
- {8 \pi^2 \over g(\mu)^2} + i \th_{SO(11)} \Bigr] }
acquires a $U(1)$ charge equal to the anomaly coefficient $C$.  The charge 
assignments for $\Lambdab$ in \anommatter\ therefore simply equal the group 
theory coefficients of the $SO(11)^2 U(1)_\V$, $SO(11)^2 U(1)_\Q$ and 
$SO(11)^2 U(1)_\R$ anomalies \ILS.

	$G$ invariance restricts the possible form of any dynamically 
generated superpotential $\Wdyn$ which can arise within a low energy 
description of the $SO(11)$ model.  For example, $\Wdyn$'s dependence upon 
$\Lambdab$ is completely fixed since it is the only field that carries 
nonvanishing R-charge.  The net numbers of spinor and vector fields 
appearing within the nonperturbative superpotential are also determined by 
$U(1)_\V$ and $U(1)_\Q$ invariance.  We thus easily find that $\Wdyn$ must 
assume the schematic forms listed in Table~1 as a function of $\Nf$.  The 
results displayed in the table suggest that the $\Nf=5$ model is analogous 
to $\Nf=\Nc$ SUSY QCD inasmuch as the R-charge assignment for $\Lambdab$ 
vanishes in this case.  As a result, no superpotential may be dynamically 
generated.  But nonperturbative effects can alter the K\"ahler potential and 
quantum mechanically constrain the matter fields.  The form such a constraint 
would have to take multiplied by a Lagrange multiplier field $X$ is shown in 
Table~1.  The $\Nf=6$ $SO(11)$ model is similarly analogous to $\Nf=\Nc+1$ 
SUSY QCD.

\vfill\eject

\parasize=1in

\begintable
$\quad \Nf \quad $ \| $\quad R(\Lambdab) \quad $ \| 
$\quad \Wdyn \quad$ \crthick
0 \| 10 \| $\bigl[\Lambda^{23} / Q^8 \bigr]^\fifth$ \nr
1 \| 8 \| $\bigl[\Lambda^{22} / V^2 Q^8 \bigr]^\quarter$ \nr
2 \| 6 \| $\bigl[\Lambda^{21} / V^4 Q^8 \bigr]^\third$ \nr
3 \| 4 \| $\bigl[\Lambda^{20} / V^6 Q^8 \bigr]^\half$ \nr
4 \| 2 \| $\Lambda^{19} / V^8 Q^8$ \nr
5 \| 0 \| $X \bigl[ V^{10} Q^8 - \Lambda^{18} \bigr]$ \nr
6 \| -2 \| $V^{12} Q^8 / \Lambda^{17}$ \endtable 
\bigskip
\centerline{Table 1: Schematic forms for dynamically generated 
superpotentials}
\bigskip

	We next need to find $SO(11)$ invariant combinations of vectors and 
spinors that act as moduli space coordinates in the low energy effective 
theory.  Equivalently, we need to determine the D-flat directions of the 
scalar potential in the microscopic theory.  Identifying independent classical 
solutions to D-flatness conditions is generally a difficult task.  However, we 
can avoid this complicated group theory exercise if we know instead the gauge 
symmetry breaking pattern realized at generic points in moduli space.  
The solution to this latter mathematical problem was worked out years ago in 
ref.~\Elashvili\ for a large class of theories including our particular 
$SO(11)$ model:
\eqn\pattern{SO(11) \> {\buildrel 32 \over \longrightarrow} \> 
	     SU(5) \> {\buildrel 11 \over \longrightarrow} \> 
	     SU(4) \> {\buildrel 11 \over \longrightarrow} \> 
	     SU(3) \> {\buildrel 11 \over \longrightarrow} \> 
	     SU(2) \> {\buildrel 11 \over \longrightarrow} \> 
	     1.}
Given this information, it is straightforward to count the number of $SO(11)$ 
singlet operators which enter into the low energy effective theory.  In 
Table~2, we display the number of parton level matter degrees of freedom as 
well as the generic unbroken color subgroup as a function of $\Nf$.  We also 
list the number of chiral superfields eaten by the superHiggs mechanism.  The 
number of independent color-singlet hadrons in the low energy theory then 
simply equals the difference between the initial and eaten matter field 
degrees of freedom.

\bigskip

\parasize=1in

\begintable
$\quad \Nf \quad $ \| Parton DOF \| Unbroken Subgroup \| Eaten DOF 
\| Hadrons \crthick 
0 \| 32 \| $SU(5)$ \| $55-24=31$ \| 1 \nr
1 \| 43 \| $SU(4)$ \| $55-15=40$ \| 3 \nr
2 \| 54 \| $SU(3)$ \| $55-8=47$ \| 7 \nr
3 \| 65 \| $SU(2)$ \| $55-3=52$ \| 13 \nr
4 \| 76 \| 1 \| 55 \| 21 \nr
5 \| 87 \| 1 \| 55 \| 32 \nr
6 \| 98 \| 1 \| 55 \| 43 \endtable 
\bigskip
\centerline{Table 2: Number of independent hadron operators}
\bigskip

	In order to figure out how to explicitly combine vector and spinor 
partons into gauge invariant hadrons, it is useful to recall some basic 
elements of $SO(11)$ group theory \refs{\Georgi{--}\Harvey}.  The tensor 
product of two 32-dimensional spinor fields decomposes into irreducible 
$SO(11)$ representations as follows:
\eqn\spinorproduct{2^5 \times 2^5 = [0]_A + [1]_S + [2]_S + [3]_A + 
[4]_A + [5]_S.}
Here $[n]$ denotes a tensor irrep with $n$ antisymmetric vector indices, 
and its ``S'' or ``A'' subscript indicates symmetry or antisymmetry under 
spinor field exchange.  Since our model contains just one spinor flavor, 
all hadrons can only involve spinor products belonging to the symmetric 
${11 \choose 1} = 11$, ${11 \choose 2}=55$ or ${11 \choose 5} = 462$ 
dimensional irreps.  We contract vector fields into these spinor combinations 
using the $SO(11)$ Gamma matrices
\eqn\Gammamatrices{
\eqalign{
\Gamma_1 &= \sigtwo \times \sigthree \times \sigthree \times \sigthree 
	\times \sigthree \cr
\Gamma_3 &= 1 \times \sigtwo \times \sigthree \times \sigthree \times 
	\sigthree \cr
\Gamma_5 &= 1 \times 1 \times \sigtwo \times \sigthree \times \sigthree \cr
\Gamma_7 &= 1 \times  1 \times 1 \times \sigtwo \times \sigthree \cr
\Gamma_9 &= 1 \times 1 \times  1 \times 1 \times \sigtwo \cr}
\qquad
\eqalign{
\Gamma_2 &= -\sigone \times \sigthree \times \sigthree \times \sigthree 
	\times \sigthree \cr
\Gamma_4 &= -1 \times \sigone \times \sigthree \times \sigthree \times 
	\sigthree \cr
\Gamma_6 &= -1 \times 1 \times \sigone \times \sigthree \times \sigthree \cr
\Gamma_8 &= -1 \times  1 \times 1 \times \sigone \times \sigthree \cr
\Gamma_{10} &= -1 \times 1 \times  1 \times 1 \times \sigone \cr}}
$$ 
\Gamma_{11} = \sigthree \times \sigthree \times \sigthree \times 
 \sigthree \times \sigthree $$
and charge conjugation matrix $C = \sigtwo \times \sigone \times \sigtwo 
\times \sigone \times \sigtwo$.  We thus form the hadrons
\eqn\hadrons{\eqalign{
L &= \bigl( Q^\T \Gamma_\mu C Q \bigr) \> \bigl( Q^\T \Gamma^\mu C Q \bigr)\cr
M^{ij} &= (V^\T)^i_\mu V^{\mu j} \cr
N^i &= Q^\T \Vslash^i C Q \cr
O^{[ij]} &= {1 \over 2!} Q^\T \Vslash^{[i} \Vslash^{j]} C Q \cr
P^{[ijklm]} &= {1 \over 5!} Q^\T \Vslash^{[i} \Vslash^j \Vslash^k \Vslash^l
	\Vslash^{m]} C Q \cr
R^{[ijklm]} &= {1 \over 5!} \e^{\mu_1 \cdots \mu_{11}} \bigl( Q^\T 
\Gamma_{\mu_1} \Gamma_{\mu_2} \Gamma_{\mu_3} \Gamma_{\mu_4} \Gamma_{\mu_5} 
C Q \bigr) \> \bigl( Q^\T \Gamma_{\mu_6} C Q \bigr) 
V^{[i}_{\mu_7} V^j_{\mu_8} V^k_{\mu_9} V^l_{\mu_{10}} V^{m]}_{\mu_{11}} \cr
T^{[ijklmn]} &= {1 \over 6!} Q^\T \Vslash^{[i} \Vslash^j \Vslash^k \Vslash^l
	\Vslash^m \Vslash^{n]} C Q.\cr}}
where Greek and Latin letters respectively denote color and flavor indices,
square brackets indicate antisymmetrization and $\Vslash^i = V^i_\mu 
\Gamma^\mu$.  Expectation values of these 
operators act as coordinates on the microscopic $SO(11)$ theory's moduli space 
of independent flat directions. 

	We can now check that the composite fields in eqn.~\hadrons\ account 
for all independent hadronic degrees of freedom within the $SO(11)$ model's 
confining phase.  In Table~3, we list the number of nonvanishing hadrons 
as a function of flavor number.  For $\Nf \le 4$, 
the $L$, $M$, $N$ and $O$ degrees of freedom sum up to the number of 
color-singlet composites entering into the low energy effective theory 
which we previously found in Table~2.  This counting works in large part due 
to the antisymmetric flavor structure of the various hadrons in \hadrons.  
When $\Nf=5$, the number of nonvanishing hadronic degrees of freedom exceeds 
the required number of composites by one.  So a single constraint must exist 
among $L$, $M$, $N$, $O$, $P$ and $R$.  This conclusion is consistent with our 
earlier finding that a relation among these fields is compatible with 
$R$-charge considerations in the $\Nf=5$ quantum theory.  Similarly, a larger 
number of independent constraints must exist in the $\Nf=6$ case in order for 
the simple counting arguments to hold.  By analogy with $\Nf=\Nc+1$ SUSY QCD, 
we expect these classical relations to persist in the quantum theory.

\bigskip

\parasize=1in

\begintable
$\quad \Nf \quad $ \| Hadrons \| $ \sp L \sp$ \| $\sp M \sp$ \| $\sp N \sp$ 
\| $\sp O \sp$ \| $\sp P \sp$ \| $\sp R \sp$ \| $\sp T \sp$ \| 
constraints \crthick 
0 \| 1  \| 1 \|    \|   \|    \|   \|   \|   \|     \cr
1 \| 3  \| 1 \| 1  \| 1 \|    \|   \|   \|   \|     \cr
2 \| 7  \| 1 \| 3  \| 2 \| 1  \|   \|   \|   \|     \cr
3 \| 13 \| 1 \| 6  \| 3 \| 3  \|   \|   \|   \|     \cr
4 \| 21 \| 1 \| 10 \| 4 \| 6  \|   \|   \|   \|     \cr
5 \| 32 \| 1 \| 15 \| 5 \| 10 \| 1 \| 1 \|   \| -1  \cr
6 \| 43 \| 1 \| 21 \| 6 \| 15 \| 6 \| 6 \| 1 \| -13 \endtable
\bigskip
\centerline{Table 3: Hadron degree of freedom count}
\bigskip

	The simple tools which we have so far utilized to analyze the $SO(11)$ 
theory restrict the possible matter content of its low energy description.  
But since these heuristic methods are clearly not rigorous, we need further 
cross checks on our conclusions regarding the $SO(11)$ model's 
vacuum structure.  We therefore examine massless parton and hadron 
contributions to global 't Hooft anomalies.  To begin, we abandon the 
anomalous global symmetry in eqn.~\anomsymgroup\ and work instead with the 
nonanomalous group
\eqn\nonanomsymgroup{
G_{\rm new}  = SO(11)_{\rm local} \times \bigl[ SU(\Nf) \times U(1)_\Y 
\times U(1)_\R \bigr]_{\rm global}.}
The generators of the new hypercharge and R-charge abelian factors are linear 
combinations of the three old $U(1)$ generators.  After assigning the partonic 
matter fields the nonanomalous charges
\eqn\nonanommatter{\eqalign{
V^{\mu i} & \sim \bigl( 11; \Nf ; -4, 1 \bigr) \cr
Q^\A & \sim \bigl( 32; 1 ; \Nf, -5/4 \bigr), \cr}}
we can readily compute the quantum numbers under $G_{\rm new}$ of 
all the composite operators in eqn.~\hadrons.

	We next compare the $SU(\Nf)^3$, $SU(\Nf)^2 U(1)_Y$, 
$SU(\Nf)^2 U(1)_R$, $U(1)_Y$, $U(1)^3_Y$, $U(1)_R$, $U(1)^3_R$, 
$U(1)^2_Y U(1)_R$, and $U(1)^2_R U(1)_Y$ global anomalies at the parton 
and hadron levels as a function of $\Nf$.  We find they precisely match 
when $\Nf=6$.  
\foot{This anomaly agreement has recently also been noted in ref.~\CSStwo.}
This nontrivial agreement is consistent with our expectation 
that the $SO(11)$ model with six vectors and one spinor confines at the 
origin of moduli space like $\Nf=\Nc+1$ SUSY QCD.   It strongly suggests that 
the low energy effective theory contains only the composite fields in 
\hadrons\ and no additional colored or colorless massless degrees of freedom.  
\foot{Anomaly matching does not shed any light upon the presence or absence 
of hadrons in the effective theory that transform as singlets under the total 
global symmetry group.  But as we shall see in the following section, such 
fields do not enter into the low energy $\Nf=6$ superpotential.  Counting 
arguments then rule out the existence of such unconstrained singlets.}
We further observe that all global anomalies match when $\Nf=5$ 
provided we include a field $X \sim \bigl( 1 ; 1 ; 0 , 2 \bigr)$ into the 
low energy spectrum.  In the quantum theory, $X$ is naturally interpreted as 
a Lagrange multiplier which enforces a single constraint.  Agreement between 
parton and hadron level anomalies occurs in other similar constrained 
$SO(\Nc)$ theories so long as Lagrange multiplier contributions are taken 
into account.  It is also important to note that global anomalies do not 
match when $\Nf \ge 7$.  The disagreement cannot be eliminated via inclusion 
of additional color-singlet composites into the low energy theory beyond those 
already listed in \hadrons.  So as in $\Nf = \Nc+2$ SUSY QCD \SeibergI, we 
interpret the anomaly mismatch as signaling the end of the $SO(11)$ model's 
confining regime and the beginning of a new dual phase.

\newsec{Low energy superpotentials}

	Having established an overall picture of the $SO(11)$ theory's 
confining phase, we are now ready to investigate its low energy structure in 
detail.  We seek to determine how nonperturbative effects in the quantum 
theory modify the classical moduli space of degenerate vacua.  One way to 
proceed is by starting with the $\Nf=0$ model and postulating that a 
superpotential consistent with the requirements of Table~1 is dynamically 
generated.  We can then try to systematically ``integrate in'' vector flavors 
and construct superpotentials in the effective theories corresponding to 
larger values of $\Nf$ \refs{\ILS,\Intriligator}.  This bottom-up approach 
unfortunately becomes intractable for $\Nf \ge 2$.  Alternatively, we can 
follow a top-down procedure in which we first deduce the form of the 
nonperturbative superpotential for $\Nf=6$ flavors.  Once $\Wdyn$ is known in 
this case, it is straightforward to methodically integrate out vector flavors 
and uncover the vacuum structure of the $SO(11)$ model for smaller 
values of $\Nf$.  We will adopt this latter approach.  

	Actually, it is technically easier to first determine the quantum 
constraint in the $\Nf=5$ theory.  Recall that $SO(11)$ breaks down to 
$SU(5)$ when the spinor field develops a nonvanishing vacuum expectation 
value.  Along the spinor flat direction, the $SO(11)$ constraint must 
reduce to the well-known $\Nf=\Nc=5$ relation $\det m - b \bar{b} = 
\Lambda_{SU(5)}^{10}$.  This requirement fixes the exact form of the $SO(11)$ 
constraint.

	In order to embed $SU(5)$ inside $SO(11)$, we first choose a set of 
24 fundamental irrep $SU(5)$ generators $t_a$ whose Cartan subalgebra members 
look like 
\eqn\Cartangens{\eqalign{
t_3 &= {1 \over \sqrt{2}} \Diag \bigl( 1,-1,0,0,0 \bigr) \cr
t_8 &= {1 \over \sqrt{6}} \Diag \bigl( 1,1,-2,0,0 \bigr) \cr
t_{15} &= {1 \over \sqrt{12}} \Diag \bigl( 1,1,1,-3,0 \bigr) \cr
t_{24} &= {1 \over \sqrt{20}} \Diag \bigl( 1,1,1,1,-4 \bigr).  \cr}}
We next form annihilation and creation operators 
\eqn\anncreatops{\eqalign{
A_j = \half \bigl( \Gamma_{2j-1} - i \Gamma_{2j} \bigr) \cr
A^\dagger_j = \half \bigl( \Gamma_{2j-1} + i \Gamma_{2j} \bigr) \cr}}
which satisfy the anticommutation relations $\{A_j, A_k \}=\{A_j^\dagger, 
A_k^\dagger \}=0$ and $\{A_j, A_k^\dagger\}=\d_{jk}$.  The $32 \times 32$ 
matrices $T_a = A_j^\dagger (t_a)_{jk} A_k$ then generate the $SU(5)$ subgroup 
of $SO(11)$ in the spinor irrep \Georgi.

	$SO(11)$ vectors and spinors break apart into $1+5+\bar{5}$ and 
$1+1+5+\bar{5}+10+\bar{10}$ under $SU(5)$.  We can explicitly see how the 
vector decomposes by inverting the relationship in eqn.~\anncreatops\ between 
the operators $A_j$ and $A_j^\dagger$ which respectively transform as $5$ and 
$\bar{5}$ under $SU(5)$ and the Gamma matrices which transform as $11$ under 
$SO(11)$:
\eqn\vectordecomp{
V^\mu = \pmatrix{ 
q_1 + \bar{q_1} \cr
i (q_1 - \bar{q_1}) \cr
q_2 + \bar{q_2} \cr
i (q_2 - \bar{q_2}) \cr
q_3 + \bar{q_3} \cr
i (q_3 - \bar{q_3}) \cr
q_4 + \bar{q_4} \cr
i (q_4 - \bar{q_4}) \cr
q_5 + \bar{q_5} \cr
i (q_5 - \bar{q_5}) \cr
\phi \cr}.}
Similarly, the $SU(5)$ irreps to which each of the spinor field's 32 elements 
belong are readily identified by acting upon $Q$ with the four Cartan 
subalgebra generators:
\eqn\spinordecomp{\eqalign{
Q^\T = \Bigl[ & \nu_R^c, \nu_L, e_L^-, e_R^+, d_L^c, d_R, u_R, u_L^c, 
	     d_L^c, d_R, u_R, u_L^c, u_R^c, u_L, d_L, d_R^c, \cr
	   & d_L^c, d_R, u_R, u_L^c, u_R^c, u_L, d_L, d_R^c, 
	     u_R^c, u_L, d_L, d_R^c, e_L^+, e_R^-, \nu_R, \nu_L^c \Bigr]. \cr}}
The names for the components of this row vector have been intentionally 
chosen to mimic familiar Standard Model and $SU(5)$ GUT nomenclature.  

	Equal nonvanishing expectation values for the first and last entries 
in the spinor field break $SO(11) \to SU(5)$ while preserving D-flatness:
\eqn\spinorvev{ \vev{Q}^\T = \bigl[ a,0,0,\cdots,0,0,a \bigr].}
This vev's dependence upon a single parameter $a$ is consistent with the 
counting argument conclusion that the $SO(11)$ model has one 
independent spinor flat direction.  Once the gauge symmetry is broken, we 
find that the $SO(11)$ hadrons decompose into the following combinations of 
$SU(5)$ mesons $m^{ij}$, baryons $b$ and $\bar{b}$, and singlets $\phi^i$:
\eqn\hadrondecomps{\eqalign{
L & \to -4 a^4 \cr
M^{(ij)} & \to 2 (m^{ij} + m^{ji}) + \phi^i \phi^j \cr
N^i  & \to 2 i a^2 \phi^i \cr
O^{[ij]} & \to -4 i a^2 (m^{ij} - m^{ji}) \cr
P & \to -32 a^2 (b+\bar{b}) + 4 i a^2 \e_{ijklm} m^{ij} m^{kl} 
  \phi^m \cr
R & \to 64 a^4 (b - \bar{b}). \cr}}
With this information in hand, we can assemble the hadrons into 
flavor singlet combinations and adjust their coefficients so as to recover 
the $SU(5)$ relation among mesons and baryons.  After a lengthy computation, 
we thus deduce the quantum constraint in the $\Nf=5$ $SO(11)$ theory:
\eqn\Wfive{\eqalign{W_{\Nf=5} &= X \Bigl\{
L^2 \det M - {1 \over 4!} \e_{i_1 i_2 i_3 i_4 i_5} \e_{j_1 j_2 j_3 j_4 j_5} 
\Bigl[ L M^{i_1 j_1} M^{i_2 j_2} M^{i_3 j_3} M^{i_4 j_4} 
N^{i_5} N^{j_5} \cr
& \qquad - 2 L M^{i_1 j_1} M^{i_2 j_2} M^{i_3 j_3} O^{i_4 j_4} O^{i_5 j_5} 
+ 6 M^{i_1 j_1} M^{i_2 j_2} N^{i_3} N^{j_3} O^{i_4 j_4} O^{i_5 j_5} \cr
& \qquad - M^{i_1 j_1} O^{i_2 j_2} O^{i_3 j_3} O^{i_4 j_4} O^{i_5 j_5} 
\Bigr] + \quarter \e_{ijklm} N^i O^{jk} O^{lm} P + L P^2 + R^2 
- \Lambda_5^{18} \Bigr\} . \cr }}

	Working in a similar fashion, we can determine the exact 
superpotential in the low energy $\Nf=6$ sigma model.  Its form 
is tightly restricted by requiring that it be smooth everywhere on the moduli 
space and that its equations of motion yield valid classical relations among 
spinor and vector fields.  Moreover, we must recover the $\Nf=5$ quantum 
constraint after giving mass to one of the vector flavors.  The unique 
superpotential which satisfies these criteria is displayed below:
\eqn\Wsix{\eqalign{W_{\Nf=6} &= {1 \over \Lambda_6^{17}} \Bigl\{
 -L^2 \det M + 
{1 \over 5!} \e_{i_1 i_2 i_3 i_4 i_5 i_6} \e_{j_1 j_2 j_3 j_4 j_5 j_6} \Bigl[ 
L M^{i_1 j_1} M^{i_2 j_2} M^{i_3 j_3} M^{i_4 j_4} M^{i_5 j_5} N^{i_6} N^{j_6}
\cr
& \qquad - {5 \over 2} 
L M^{i_1 j_1} M^{i_2 j_2} M^{i_3 j_3} M^{i_4 j_4} O^{i_5 j_5} O^{i_6 j_6} 
+ 10 M^{i_1 j_1} M^{i_2 j_2} M^{i_3 j_3} N^{i_4} N^{j_4} O^{i_5 j_5} 
O^{i_6 j_6} \cr
& \qquad - {5 \over 2} 
M^{i_1 j_1} M^{i_2 j_2} O^{i_3 j_3} O^{i_4 j_4} O^{i_5 j_5} O^{i_6 j_6} 
\Bigr] - R_i M^{ij} R_j - L P_i M^{ij} P_j \cr
& \qquad  + 2 i P_i O^{ij} R_j + (N^i P_i)^2 
+ \quarter P_i M^{ij} \e_{jklmno} N^k O^{lm} O^{no} \cr
& \qquad  + L T^2 + 2 i N^i R_i T - 2 ({\rm Pf} \, O) T \Bigr\}. }}
We note that all combinations of $SO(11)$ hadrons consistent with symmetry and 
smoothness considerations enter into $W_{\Nf=6}$ with nonvanishing 
coefficients.  Although this result may seem natural, other theories analogous 
to $\Nf=\Nc+1$ SUSY QCD are known to have zero coefficients for some 
{\it a priori} allowed terms in their superpotentials 
\refs{\ChoKraus,\CSSone}.  So while it is relatively easy to figure out the 
basic polynomial form of the numerator in \Wsix\ as it is in many similar 
models \CSStwo, determining the numerical values for each term's coefficient 
requires a detailed computation.  

	Once the ground state structures of the $\Nf=5$ and $\Nf=6$ $SO(11)$ 
theories are known, it is straightforward to flow down to models with fewer 
vector fields.  We simply add a tree level mass term $W_{\rm tree} = 
\mu_{ij} M^{ij}$ to eqn.~\Wfive, rotate the meson field $M^{ij}$ into diagonal 
form via a flavor transformation and then integrate out heavy vector flavors.  
Since the modified quantum theory with the meson mass term contains no sources 
which transform under the $SU(\Nf)$ flavor group like $N^i$ or $O^{[ij]}$, 
the expectation values of those fields containing heavy vectors must vanish.  
After systematically removing heavy vector flavors, we find the following 
tower of dynamically generated $SO(11)$ superpotentials:
\foot{All flavor indices on $M^{ij}$, $N^i$ and $O^{ij}$ in $W_{(\Nf=4)}$, 
$W_{(\Nf=3)}$ and $W_{(\Nf=2)}$ are implicitly contracted with $SU(\Nf)$ 
epsilon symbols so as to form flavor singlet combinations.}
$$ W_{(\Nf=4)} = {\Lambda_4^{19} \over L^2 M^4 - 4 L M^3 N^2 
    + 6 L M^2 O^2 - 12 M N^2 O^2 + O^4 } $$
\centerline{$\downarrow$}
$$ W_{(\Nf=3)} = 2 \Bigl[ {\Lambda_3^{20} \over 
L^2 M^3 - 3 L M^2 N^2 + 3 L M O^2 - 3 N^2 O^2 } \Bigr]^{\half} $$ 
\eqn\Wzerotofour{\downarrow}
$$ W_{(\Nf=2)} = 3 \Bigl[ {\Lambda_2^{21} \over 
L^2 M^2 - 2 L M N^2 + L O^2 } \Bigr]^{\third} $$
\centerline{$\downarrow$}
$$ W_{(\Nf=1)} = 4 \Bigl[ {\Lambda_1^{22} \over L^2 M - L N^2} 
\Bigr]^{\quarter} $$
\centerline{$\downarrow$}
$$ W_{(\Nf=0)} = 5 \Bigl[ {\Lambda_0^{23} \over L^2} \Bigr]^{1 \over 5}. $$ 
The strong interaction scales are related by requiring gauge coupling 
continuity across heavy vector thresholds:
\eqn\scalerelns{\eqalign{
\Lambda_0^{23} &= \mu_{11} \Lambda_1^{22} = \mu_{11} \mu_{22} \Lambda_2^{21} 
= \mu_{11} \mu_{22} \mu_{33} \Lambda_3^{20} 
= \mu_{11} \mu_{22} \mu_{33} \mu_{44} \Lambda_4^{19} \cr
& = \mu_{11} \mu_{22} \mu_{33} \mu_{44} \mu_{55} \Lambda_5^{18}
  = \mu_{11} \mu_{22} \mu_{33} \mu_{44} \mu_{55} \mu_{66} \Lambda_6^{17}.}}
As a check, one can verify that these $\Nf < 5$ $SO(11)$ expressions
properly reduce to their $SU(5)$ descendants 
\eqn\sufiveWs{W_{\Nf < 5} = (\Nc-\Nf) \Bigl[ {\Lambda^{15-\Nf}_{SU(5)} \over 
\det m} \Bigr]^{1 \over 5-\Nf} }
along the spinor flat direction.

	The nonperturbative superpotentials in \Wzerotofour\ lift the 
classical vacuum degeneracy and generate runaway scalar potentials.  The 
$\Nf \le 4$ $SO(11)$ quantum theories can be stabilized by adding tree 
level superpotentials that increase along all D-flat directions.  If 
$W_{\rm tree}$ preserves some global symmetry which is spontaneously broken 
in the true ground state, the chiral $SO(11)$ model should dynamically break 
supersymmetry \ADS.  This condition on $W_{\rm tree}$ cannot 
be satisfied in the $\Nf=0$ theory, for all polynomial functions of the spinor 
field $Q$ break the global $U(1)_\R$ symmetry.  Witten index arguments then
suggest that SUSY remains unbroken in models with additional vector fields.
Nevertheless, supersymmetry may be broken via other schemes such as 
coupling singlets to all the hadrons in the $\Nf=5$ quantum constraint \IT.
So whether SUSY can be dynamically broken in this $SO(11)$ theory 
remains an interesting open question. 

\newsec{Conclusion}

	In this paper, we have investigated the low energy behavior of 
$SO(11)$ models containing $\Nf \le 6$ vectors and one spinor matter field. 
The tricks and tools which we used to analyze this particular theory can be 
profitably applied to the study of many other $\CN=1$ models.  Knowing the 
pattern of gauge symmetry breaking at generic points in moduli space is 
especially valuable.  Indeed, the confining phase structure of the $SO(11)$ 
theory is essentially fixed by $\Nc=5$ SUSY QCD along its spinor flat 
direction.  Other $SO(\Nc)$ models of similar type are likewise restricted
\refs{\Pouliot{--}\Kawano}.

	It would be highly desirable to extend our understanding of the 
strongly interacting $SO(11)$ model into its dual phase.  Our present 
confining phase results provide helpful clues in the search for a weakly 
coupled dual.  In particular, the superpotential in eqn.~\Wsix\ must be 
recovered from any magnetic dual to the $SO(11)$ electric theory when the 
number of vector flavors is reduced down to six.  Our primary incentive for
explicitly calculating $W_{\Nf=6}$ was in fact to determine which 
nonvanishing terms must be reproduced by a magnetic theory.  The complex 
superpotential expression in \Wsix\ suggests the dual is not simple.  

	Given that the spinor flat direction played a central role in our 
analysis of the $SO(11)$ theory's confining phase, we naturally want to 
exploit it for studying the nonabelian Coulomb phase as well.  Motivated by 
Seiberg's dual to $\Nc=5$ SUSY QCD \SeibergII, we speculate that the magnetic 
gauge group $\tilde{G}$ contains $SU(\Nf-5)$ as a subgroup.  Of course, 
other duals beside Seiberg's could exist for $SU(5)$ gauge theory with 
$\Nf \ge 7$ quark flavors. So $\tilde{G}$ need not resemble $SU(\Nf-5)$ at 
all.

	In closing, we mention that a weakly coupled magnetic description 
of the electric $SO(11)$ theory would yield several other dual pairs 
as interesting by-products.  For example after Higgsing the $SO(11)$ gauge
group, one should find duals to $SO(6) \simeq SU(4)$ and $SO(5) \simeq Sp(4)$ 
models with $\Nf=4$ quark flavors and various numbers of antisymmetric fields.  
These special cases might shed light upon more general $SU(2\Nc)$ and 
$Sp(2\Nc)$ models with fundamental and antisymmetric matter.  Finding weakly 
coupled duals to these theories with zero tree level superpotential remains an 
unsolved and challenging problem.

\bigskip\bigskip\bigskip
\centerline{{\bf Acknowledgments}}
\bigskip

        It is a pleasure to thank Howard Georgi and Matt Strassler for 
sharing their insights.  I would also like to especially thank Per Kraus for 
collaborating at the beginning of this work and for numerous helpful 
discussions throughout this study.  Finally, I am grateful to Csaba Cs\`aki 
and Witold Skiba for their comments on the manuscript.  

\vfill\eject

\listrefs
\bye